\documentstyle[12pt,a4]{article}
\begin{document}
\baselineskip=19 true pt

\title{Josephson vortex Bloch oscillations:\\
single pair tunneling effect}
\author{R.G.~Mints$^\ast$ and I.B.~Snapiro$^\dagger$\\
\\
$^\ast$School of Physics and Astronomy, \\
Raymond and Beverly Sacler Faculty of Exact Sciences, \\
Tel--Aviv University, 69978 Ramat--Aviv, Israel\\
\\
$^\dagger$Physics Department, \\
Technion--Israel Institute of Technology, Haifa 32000, Israel
}
\date{}
\maketitle
\vskip  2 true cm
We consider the Josephson vortex motion in a long
one--dimensional Josephson junction in a thin film. We show
that this Josephson vortex is similar to a mesoscopic
capacitor. We demonstrate that a single Cooper pair
tunneling results in nonlinear Bloch--type oscillations of
a Josephson vortex in a current-biased Josephson junction.
We find the frequency and the amplitude of this motion.

\vfill
\noindent {PASC numbers: 74.60. Ec, 74.60. Ge} \par
\eject
Single electron and Cooper pair tunneling remain a
fundamental problem in normal metal and superconductivity
physics. Under detail experimental and theoretical
study is electron transport through tunnel junctions in
mesoscopic systems \cite{b1,b2}. A particular attention
has been given also to the classical and quantum dynamics
of Josephson vortices in Josephson junctions and arrays of
Josephson junctions. Very fast classical motion of Josephson
vortices with velocities approaching the Swihart velocity
was treated experimentally in annular Josephson tunnel
junctions \cite{b3,b4}. The ballistic motion of vortices in
two-dimensional arrays of Josephson junctions is under
thorough theoretical and experimental study \cite{b5,b6}.
It was shown that the spin waves excitation results in an
intrinsic friction when a vortex is moving in an array of
Josephson junctions \cite{b7,b8,b9,b10,b6}. Vortices
tunneling is treated theoretically and experimentally in
two-dimensional arrays of Josephson junctions
\cite{b11,b12,b6}.
\par
In this Letter we consider the effect of a quantum single
Cooper pair tunneling on the classical motion of a Josephson
vortex in a long Josephson junction. We point out that a
Josephson vortex in a thin film is similar to a mesoscopic
capacitor. We show that at sufficiently low temperatures it
results in nonlinear Bloch--type oscillations in a
current--biased Josephson junction. We find the amplitude
and frequency of this motion.
\par
To develop the analogy between a Josephson vortex and a
mesoscopic capacitor let us consider a superconducting film
with a Josephson junction along the $x$ axis (see Fig.~1).
The motion of a Josephson vortex along this Josephson
junction induces voltage $V(x,t)$. The value of the voltage
drop across the Josephson junction is equal to \cite{b13}
\begin{equation}
V={\hbar\over 2e}{\partial\varphi\over\partial t},
\label{e1}
\end{equation}
where $\varphi (x,t)$ is the phase difference. This voltage
results in a certain charge $Q$ moving simultaneously with
the Josephson vortex. As $\varphi (x,t)=\varphi (x-vt)$ and
$\varphi (+\infty)-\varphi (-\infty)=2\pi$ the value of $Q$
takes the explicit form
\begin{equation}
Q=\int\limits^\infty_{-\infty}\rho (x-vt)\,dx=
{\hbar\over 2e}\,dC\int\limits^\infty_{-\infty}
{\partial\varphi\over\partial t}\,dx=
{\pi\hbar\over e}\,dCv,
\label{e2}
\end{equation}
where $d$ is the film thickness, $C$ is the specific
capacity of the Josephson junction, and $v$ is the
Josephson vortex velocity. The charge density $\rho (x-vt)$
is distributed in the cross--section of the Josephson
junction over an area of the order of $dl$, where $l$ is
the characteristic space scale of the phase difference
distribution, {\it i.e.}, the size of the Josephson vortex.
\par
Thus, a moving Josephson vortex is similar to a charged
mesoscopic capacitor with a certain effective capacity
$C_{\rm eff}$. The value of $C_{\rm eff}$ is proportional to
the area $dl$, where the charge $Q$ is localized. Mesoscopic
effects affect the motion of a Josephson vortex if the
effective capacity $C_{\rm eff}\propto dl$ is small,
{\it i.e.}, in the case of small $d$ and $l$.
\par
Let us now consider a Josephson vortex in a thin film
meaning that $d\ll\lambda$, where $\lambda$ is the London
penetration depth. The space and time dependencies of the
phase difference $\varphi (x,t)$ determine the statics and
dynamics of a Josephson vortex. To find the dependence
$\varphi (x,t)$ we use the continuity equation
${\rm div}\,{\bf j}=0$, where ${\bf j}$ is the current
density. We treat here the case of zero dissipation,
{\it i.e.}, the case of an ideal Josephson junction and
calculate the current density taking into account the
superconducting, tunneling and displacement currents. We
neglect the resistive (leakage and quasi-particle tunneling)
currents as they can be minimized by the Josephson junction
design and by lowering the temperature. An additional friction
force arise as a result of Josephson vortex Cherenkov radiation
of small amplitude electromagnetic waves propagating along
the tunnel junction \cite{b14}. The contribution of this
radiation friction force becomes important only when the
Josephson vortex velocity is approaching the Swihart velocity.
Note, that the dissipation due to Josephson vortex Cherenkov
radiation in a long Josephson junction is similar to the
dissipation due to spin waves excitation in Josephson junction
arrays \cite{b7,b8,b9,b10,b6}.
\par
In the thin film limit the density of the superconducting
current and magnetic field in the superconductor decrease
with the characteristic space scale \cite{b15}
\begin{equation}
\lambda_{\rm eff}={\lambda^2\over d}\gg\lambda.
\label{e3}
\end{equation}
\par
To emphasize on the mesoscopic effects we consider the case
of a small size Josephson vortex, {\it i.e.}, we assume that
$l\ll \lambda_{\rm eff}$. Then the relation between the
superconducting current density and the phase difference
becomes nonlocal and the continuity equation leads to a
nonlinear integro--differential equation determining the
value of $\varphi (x,t)$ \cite{b16,b17}. In particular, in the case
of $d\ll\lambda$ and $l\ll \lambda_{\rm eff}$ this equation
reads \cite{b18}
\begin{equation}
\sin\varphi + {1\over\omega^2_p}\,
{\partial^2\varphi\over\partial t^2}=
{l_J\over\pi}\,\int\limits_{-\infty}^\infty
{dy'\over y'-y}\, {\partial\varphi\over\partial y'},
\label{e4}
\end{equation}
where
\begin{equation}
\omega_p=\Bigl({2ej_c\over C\hbar}\Bigr)^{1/2},
\label{e5}
\end{equation}
is the plasma frequency,
\begin{equation}
l_J={c\Phi_0\over 16\pi^2\lambda^2j_c},
\label{e6}
\end{equation}
$\Phi_0$ is the flux quantum, and $j_c$ is the Josephson
critical current density.
\par
Let us first consider a stationary Josephson vortex. In this
case  $v=0$ and the space distribution of the phase
difference $\varphi (x)$ is given by the exact solution of
Eq.~(\ref{e4}) \cite{b16,b17}
\begin{equation}
\varphi (x)=2\arctan \Bigl({x\over l_J}\Bigr)+\pi,
\label{e7}
\end{equation}
If the velocity $v$ of a Josephson vortex is small,
{\it i.e.}, $v\ll\omega_p l_J$ the solution of Eq.~(\ref{e4}) is
given by Eq.~(\ref{e7}), where instead of $x$ we substitute $x-vt$.
In particular, it follows from Eq.~(\ref{e7}) that the size of a
Josephson vortex moving with a velocity $v\ll\omega_p l_J$
is of the order of $l_J$.
\par
The motion of a Josephson vortex induces an electrical
field localized inside the Josephson junction. Using Eq.~(\ref{e1})
we find that the energy of this electrical field is equal to
\begin{equation}
{\cal E}_v={\hbar^2Cd\over 8e^2}\,
\int\limits_{-\infty}^\infty
\Bigl({\partial\varphi\over\partial t}\Bigr)^2 dx=
{\hbar^2Cdv^2\over 8e^2}\,\int\limits_{-\infty}^\infty
\Bigl({\partial\varphi\over\partial x}\Bigr)^2 dx.
\label{e8}
\end{equation}
It follows from Eq.~(\ref{e8}) that ${\cal E}_v\propto v^2$.
Combining Eqs. (\ref{e8}) and (\ref{e2}) we find that at the same time
${\cal E}_v\propto Q^2$. We can thus introduce the
Josephson vortex mass $M$ and effective capacity
$C_{\rm eff}$ presenting ${\cal E}_v$ as
\begin{equation}
{\cal E}_v={Mv^2\over 2}={Q^2\over 2C_{\rm eff}},
\label{e9}
\end{equation}
where
\begin{equation}
M={\pi\hbar^2Cd\over 2e^2l_J},
\label{e10}
\end{equation}
and
\begin{equation}
C_{\rm eff}=2\pi Cdl_J.
\label{e11}
\end{equation}
\par
Let us now consider the motion of a Josephson vortex along
a Josephson junction. We treat the case when the driving
force is the Lorentz force $F_L$ resulting from a bias
current across the Josephson junction. Then the equation of
motion reads
\begin{equation}
M{dv\over dt}={\Phi_0jd\over c}.
\label{e12}
\end{equation}
where $j$ is the bias current density. Using Eq.~(\ref{e2}) we
rewrite Eq.~(\ref{e13}) as
\begin{equation}
{dQ\over dt}=2\pi jdl_J.
\label{e13}
\end{equation}
It follows from Eqs.~(\ref{e12}) and (\ref{e13}) that the bias current
results in the Josephson vortex acceleration and charging.
\par
Let us now treat a single Cooper pair crossing a Josephson
junction. This elementary recharging process ($Q\to Q-2e$)
is changing the electrical field energy of a Josephson
vortex by $\Delta{\cal E}_v={\cal E}_v(Q-2e)-{\cal E}_v(Q)$.
A single Cooper pair tunneling can happen if the energy
difference $\Delta{\cal E}_v$ is equal to zero. It follows
from Eq.~(\ref{e9}) that $\Delta{\cal E}_v=0$ when the charge
$Q=e$. The value of $Q$ is equal to the electron charge
$e$ at a certain Josephson vortex velocity $v_m$. It follows
from Eq.~(\ref{e2}) that
\begin{equation}
v_m={e^2\over\pi\hbar}{1\over Cd}.
\label{e14}
\end{equation}
\par
This reasoning is true if the elementary charging energy of
a Josephson vortex ${\cal E}_v(e)$ is larger than the scale
of the thermal fluctuations $k_BT$, {\it i.e.}, if
\begin{equation}
{e^2\over 2C_{\rm eff}}={e^2\over 4\pi Cdl_J}\gg k_BT,
\label{e15}
\end{equation}
where T is the temperature and $k_B$ is the Boltzmann
constant. Note, that the inequality given by
Eq.~(\ref{e15}) restricts the film thickness to
$d\ll e^2/4\pi k_BTCl_J$.
\par
Thus, a single Cooper pair tunneling is changing the charge
of a moving Josephson vortex from $e$ to $-e$. It follows
from Eq.~(\ref{e2}) that simultaneously this tunneling event is
changing the velocity of a Josephson vortex from $v_m$ to
$-v_m$. It means that in the mainframe of a classical
mechanics approach to the motion of a Josephson vortex a
single Cooper pair tunneling reveals as an elastic impact.
The effect of this impact is as follows. The Lorentz force
$F_L$ is increasing the velocity of a Josephson vortex $v$
with a constant acceleration $a_v=F_L/M$. At the moment when
the value of $v$ becomes equal to $v_m$ an elastic impact
(a single Cooper pair tunneling event) happens changing
$v_m$ to $-v_m$. As the velocity of a Josephson vortex is
continuing to increase with the rate given by $a_v$ the
process repeats itself periodically. It follows from
Eqs.~(\ref{e12}) and (\ref{e13}) that the frequency $\omega$ and the
amplitude $l_m$ of this periodic Josephson vortex motion
are given by
\begin{equation}
\omega={2\pi^2jdl_J\over e},
\label{e16}
\end{equation}
\begin{equation}
l_m={e^3\over 4\pi^2\hbar Cd^2jl_J}.
\label{e17}
\end{equation}
The dependence of the Josephson vortex velocity on time is
shown in Fig.~2.
\par
To estimate the values of $\omega$ and $l_m$ we substitute
in Eq.~(\ref{e17}) the expression for $l_J$ given by Eq.~(\ref{e6}) and
the specific capacity $C$ in the form
\begin{equation}
C={\epsilon\over 4\pi d_0},
\label{e18}
\end{equation}
where $\epsilon$ is the dielectric constant and $d_0$ is
the distance between the banks of the Josephson junction.
It results in the formulae
\begin{equation}
\omega={\pi\over 8}\,{\hbar c\over e^2}\,{j\over j_c}\,
{cd\over\lambda^2},\qquad {\rm and}\qquad
l_m={16\over\epsilon}\,\Bigl({e^2\over\hbar c}\Bigr)^2\,
\Bigl({\lambda\over d}\Bigr)^2\,{j_c\over j}\,d_0.
\label{e19}
\end{equation}
Using the data $\lambda\approx 3\times 10^{-5} cm$,
$d\approx 10^{-6} cm$, $d_0\approx 10^{-8} cm$, and
$\epsilon\approx 5$, we find an estimation for the values
of $\omega$ and $l_m$ in the form
\begin{equation}
\omega\approx 2\times 10^{15}\,{j\over j_c}\,s^{-1},
\qquad{\rm and}\qquad
l_m\approx 1.5\times 10^{-9}\,{j_c\over j}\,cm.
\label{e20}
\end{equation}
It follows from Eqs.~(\ref{e19}) and (\ref{e20}) that if
$j/j_c\approx 10^{-8}$ then
$\omega\approx 2\times 10^7\,s^{-1}$
and $l_m\approx 0.15\,cm$. Note, that for the same data as
above and the critical current density
$j_c\approx 10^5\,A/cm^2$ we estimate
$l_J\approx 1.5\times 10^{-4}\,cm$,
$C_{\rm eff}\approx 7.5\times 10^{-3}\,cm$ and the
characteristic temperature
\begin{equation}
{e^2\over 2C_{\rm eff}k_B}\approx 0.1\,K.
\label{e21}
\end{equation}
\par
To summarize, we show that a Josephson vortex in a thin
film is similar to a mesoscopic capacitor. A single Cooper
pair tunneling results in Josephson vortex nonlinear
Bloch--type oscillations in a current--biased Josephson
junction. We find the frequency and the amplitude of this
motion.
\par
We are grateful to Y.~Gefen, D.E.~Khmelnitskii and
M.~Tinkham for useful discussions. This work was
supported in part by the Foundation Raschi.
\vfill\eject

\vfill\eject

\ \ \ \ \ \par
\vskip 1 true cm
\centerline {CAPTIONS}
\vskip 1 true cm
\begin{description}
\item[Fig.~1.] A superconducting film with a Josephson
junction (thick line).
\item[Fig.~2.] The dependence of the Josephson vortex
velocity on time (arbitrary units).
\end{description}
\vfill

\begin{thebibliography}{22}
\bibitem{b1} D.V.~Averin and K.K.~Likharev, in {\it
Mesoscopic Phenomena in Solids}, edited by B.~Altshuler
{\it et al.} (Elsevier, Amsterdam, 1991), p.~173.
\bibitem{b2} {\it Single Charge Tunneling}, edited by
H.~Grabert and M.H.~Devoret (Plenum, New York, 1992).
\bibitem{b3} A.V.~Ustinov, T.~Doderer, B.~Mayer, R.P.~Huebener
and V.A.~Oboznov, Europhys.~Lett. {\bf 19}, 63 (1992).
\bibitem{b4}  A.V.~Ustinov, T.~Doderer, R.P.~Huebener,
N.F.~Pedersen, B.~Mayer and V.A.~Oboznov, Phys. Rev. Lett. {\bf 69},
1815 (1992).
\bibitem{b5} H.S.J. van der Zant, F.C.~Fritschy, T.P.~Orlando
and J.E.~Mooij, Europhys. Lett. {\bf 18}, 343 (1992).
\bibitem{b6} R.~Fazio, A.~van~Otterlo and G.~Sch\"{o}n,
Europhys.~Lett. {\bf 25}, 453 (1994).
\bibitem{b7} P.~Bobbert, Phys. Rev. {\bf B 45}, 7540 (1991).
\bibitem{b8} U.~Eckern and E.B.~Sonin, Phys. Rev. {\bf B 47},
505 (1993).
\bibitem{b9} U.~Geigenm\"{u}ller, C.J.~Lobb and C.B.~Whan,
Phys. Rev. {\bf B 47}, 348 (1993).
\bibitem{b10} H.S.J. van der Zant, F.C.~Fritschy, T.P.~Orlando,
J.E.~Mooij, Phys. Rev. {\bf B 47}, 295 (1993).
\bibitem{b11}  H.S.J. van der Zant, F.C.~Fritschy, T.P.~Orlando,
J.E.~Mooij,  Phys. Rev. Lett. {\bf 66}, 2531 (1991).
\bibitem{b12} T.~Tighe, A.T.~Johnson and M.~Tinkham,
Phys. Rev. {\bf B 44}, 10286 (1991).
\bibitem{b13} A.~Barone and G.~Paterno, {\it Physics and
Applications of the Josephson Effect} (Wiley, New York,
1982).
\bibitem{b14} R.G.~Mints and I.B.~Snapiro, Phys.~Rev.
{\bf B },\ \  (1994).
\bibitem{b15} J.~Pearl, Appl. Phys. Lett. {\bf 5}, 65 (1964).
\bibitem{b16} A.~Gurevich, Phys. Rev. {\bf B 46}, 3187 (1992).
\bibitem{b17} R.G.~Mints and I.B.~Snapiro, Physica {\bf A 200},
426 (1993).
\bibitem{b18} R.G.~Mints and I.B.~Snapiro, Phys. Rev.
{\bf B 49}, 6188 (1994).

\end{thebibliography}
\end{document}